\documentclass[english,american,aip,jcp,reprint,superscriptaddress]{revtex4-1}
\usepackage[T1]{fontenc}
\usepackage[utf8]{inputenc}
\usepackage{xcolor}
\usepackage{babel}
\usepackage{amsmath}
\usepackage{amssymb}
\usepackage{graphicx}
\PassOptionsToPackage{normalem}{ulem}
\usepackage{ulem}
\usepackage[unicode=true,
 bookmarks=true,bookmarksnumbered=false,bookmarksopen=false,
 breaklinks=true,pdfborder={0 0 0},pdfborderstyle={},backref=false,colorlinks=true]
 {hyperref}
\hypersetup{pdftitle={Electronic and optical properties of hydroxyapatite from first-principles: impact of the exchange-correlation treatment},
 pdfauthor={Leon A. Avakyan et al.},
 allcolors=blue}
\usepackage{breakurl}

\makeatletter

\providecommand{\tabularnewline}{\\}
\providecolor{lyxadded}{rgb}{0,0,1}
\providecolor{lyxdeleted}{rgb}{1,0,0}

\DeclareRobustCommand{\lyxsout}[1]{\ifx\\#1\else\sout{#1}\fi}

\makeatother

\begin{document}

\title{Optoelectronics and defect levels in hydroxyapatite by first-principles }

\author{Leon A. Avakyan}
\email{laavakyan@sfedu.ru}

\selectlanguage{american}%

\affiliation{Department of Physics and I3N, University of Aveiro, Campus Santiago,
3810-193 Aveiro, Portugal}

\affiliation{Physics Faculty, Southern Federal University, Zorge street 5, Rostov-on-Don
344090, Russian Federation}

\author{Ekaterina V. Paramonova}

\affiliation{Institute of Mathematical Problems of Biology, Keldysh Institute
of Applied Mathematics, Russian Academy of Sciences, Vitkevicha street
1, Pushchino, 142290, Moscow region, Russian Federation}

\author{José Coutinho}

\affiliation{Department of Physics and I3N, University of Aveiro, Campus Santiago,
3810-193 Aveiro, Portugal}

\author{Sven Öberg}

\affiliation{Department of Engineering Sciences and Mathematics, Luleå University
of Technology, SE-97187 Luleå, Sweden}

\author{Vladimir S. Bystrov}

\affiliation{Institute of Mathematical Problems of Biology, Keldysh Institute
of Applied Mathematics, Russian Academy of Sciences, Vitkevicha street
1, Pushchino, 142290, Moscow region, Russian Federation}

\author{Lusegen A. Bugaev}

\affiliation{Physics Faculty, Southern Federal University, Zorge street 5, Rostov-on-Don
344090, Russian Federation}
\begin{abstract}
Hydroxyapatite (HAp) is an important component of mammal bones and
teeth, being widely used in prosthetic implants. Despite the importance
of HAp in medicine, several promising applications involving this
material (\emph{e.g.} in photo-catalysis), depend on how well we understand
its fundamental properties. Among the ones that are either unknown
or not known accurately we have the electronic band structure and
all that relates to it, including the band gap width. We employ state-of-the-art
methodologies, including density hybrid-functional theory and many-body
perturbation theory within the GW approximation, to look at the optoelectronic
properties of HAp. These methods are also applied to the calculation
of defect levels. We find that the use of a mix of (semi-)local and
exact exchange in the exchange-correlation functional, brings a drastic
improvement to the band structure. Important side-effects include
improvements in the description of dielectric and optical properties,
not only involving conduction band (excited) states, but also the
valence. We find that the highly dispersive conduction band bottom
of HAp originates from anti-bonding $\sigma^{*}$ states along the
$\cdots\textrm{OH-OH-}\cdots$ infinite chain, suggesting the formation
of a conductive 1D-ice phase. The choice of the exchange-correlation
treatment to the calculation of defect levels was also investigated
by using the OH-vacancy as testing-model. We find that donor and acceptor
transitions obtained within semi-local DFT differ from those of hybrid-DFT
by almost 2~eV. Such a large discrepancy emphasizes the importance
of using a high-quality description of the electron-electron interactions
in the calculation of electronic and optical transitions of defects
in HAp. \emph{Published in Journal of Applied Physics }\textbf{\emph{148}}\emph{,
154706 (2018).} \texttt{\textbf{\textcolor{blue}{https://doi.org/10.1063/1.5025329}}}
\end{abstract}
\maketitle

\section{Introduction}

Ceramics based on calcium apatites are important materials in health
care, biology, ecology and catalysis.\cite{Bystrov2017,Duminis2017}
Among the apatites, hydroxyapatite (HAp) is singled out as the main
component of vertebrate bones and teeth. Due to its innate bio-activity,
HAp became a widely used biomaterial in medicine, with applications
on bone and dental implants (for instance as filler and coating material),\cite{Ratner2004book,Epple2010,Bystrov2011}
as well as an absorbent for liquid chromatography.\cite{Nishikawa2002,Bystrov2015} 

Much of the emerging interest on HAp and related materials stems from
their photo-catalytic and ferroelectric properties.\cite{Lang2013,hu2017,Piccirillo2017}
Ultra-violet (UV) assisted filtration and decomposition of pharmaceutical
pollutants, dyes and bacteria inactivation have been demonstrated.\cite{Nishikawa2004,Nishikawa2005,Ozeki2007,Reddy2007,Shariffuddin2013,MarquezBrazon2016,Anirudhan2017}
For these applications, modified HAp-based materials or composites,
such as Ti-doped HAp,\cite{Piccirillo2017} TiO$_{2}$-supported HAp,\cite{Ozeki2007,MarquezBrazon2016}
graphene-oxide supported HAp,\cite{Anirudhan2017} as well as thermally
modified material,\cite{Nishikawa2004,Reddy2007,Shariffuddin2013}
are the most promising.

The origin of the photocatalytic activity of HAp has been attributed
to the production of superoxide O$_{2}^{\bullet-}$ radicals, following
the absorption of UV light by near-surface vacancies and subsequent
electron transfer to oxygen in the atmosphere.\cite{Nishikawa2002,Nishikawa2004_lett,Nishikawa2004}
These observations were recently supported by some of us by means
of density functional calculations.\cite{Bystrov2016}

In order to have an accurate picture of the photocatalytic mechanisms
taking place, it is crucial that we are provided with accurate figures
for the fundamental properties of the material, as well as methodologies
to estimate them. These include, the crystalline structure, phonon
dispersion, electronic band structure, dielectric response, electronic
band gap, among others. Due to poor sample quality, most of these
quantities are associated with large error bars. For instance, the
measured width of the forbidden electronic gap ($E_{\mathrm{g}}$)
ranges from above 6~eV down to 3.95~eV.\cite{Rosenman2007,Tsukada2011}
Analogously, and despite much progress on the theoretical side, density
functional theory (DFT) based on a local or semi-local description
of the exchange-correlation interactions between electrons, give $E_{\mathrm{g}}$
values between 4.5~eV and 5.4~eV, also depending on the particular
choice of basis type for describing the valence states.\cite{Calderin2003,Rulis2004,Matsunaga2007,Slepko2011,Bystrov2015}
As prudently noted by the authors of Ref.~\onlinecite{Slepko2011},
the inherent insufficiencies of a (semi-)local exchange-correlation
approximation severely underestimate the band gap of insulators, and
therefore, it is likely that for HAp the true value of $E_{\mathrm{g}}$
lies well above 5.5~eV.

Defect-free HAp is transparent to visible light and the above arguments
suggest that electronic excitations are achievable only under middle-
or far-ultraviolet illumination, most probably with $h\nu>6$~eV.
The report of Nishikawa and co-workers\cite{Nishikawa2002} on the
photocatalytic activity using a light source of $h\nu=4.88$~eV ($\lambda=254$~nm),
implies that behind this effect we should have a transition involving
at least one bound state located in the gap, most probably associated
with defects. Now the question is: Are we dealing with vacancies?
Or should that be caused by other point defects, impurities, grain
boundaries or even surfaces states? In previous studies several point
defects in HAp were inspected by using the local density approximation
(LDA) to the exchange-correlation potential. It was found that an
oxygen vacancy in the phosphate (PO$_{4}$) unit introduces a fully
occupied one-electron level in the lower half of the gap.\cite{Bystrov2016,Bystrov2017}
The gap of the defective system was claimed to be compatible with
the photocatalytic threshold energy. However, the lack of physical
significance of the Kohn-Sham eigenvalues along with large self-interaction
errors from the LDA,\cite{Martin2016} makes any assignment of an
optical transition to a defect based on its Kohn-Sham band structure,
highly speculative.

We have therefore to (i) employ state-of-the-art techniques to deal
with the exchange-correlation problem, and to (ii) use physically
sound methodologies to obtain defect-related transitions. The first
issue has been successfully addressed in wide-gap materials, by admixing
a fraction of exact exchange with the local or semi-local exchange.
The exact exchange term is obtained from Hartree–Fock theory, it is
fully non-local and free from electronic self-interactions. The resulting
functionals have been termed \emph{hybrid} and in general they improve
molecular ionization energies and affinities, as well as the description
of the band structure of solids, particularly of wide-gap insulators.\cite{Lee1988,Becke1993,Stephens1994,Adamo1999,Heyd2003,Krukau2006}
The second issue must be addressed by calculating defect transitions
solely based on the many-body energy obtained from the density functional.
Two methods have been widely used, namely from the defect formation
energy as function of its charge state,\cite{Qian1988} or by comparing
defect ionization energies and electron affinities (still ground state
properties) with analogous quantities calculated for defects with
well characterized levels.\cite{Resende1999}

We also note that since the underestimation of the gap affects the
one-electron contribution to the total energy, it will also affect
the calculation of quantities that strongly depend on the quality
of the band structure, like migration barriers, work-functions, dielectric
response, among many others. For instance, Corno \emph{et~al.}\cite{Corno2006}
investigated the structural and vibrational properties of HAp using
the B3LYP hybrid exchange functional.\cite{Becke1993} Not surprisingly,
they reported $E_{\mathrm{g}}=7.9$~eV, about 2.5~eV wider than
the LDA and the generalized gradient approximated (GGA) results.

The availability of alternative hybrid exchange-correlation functionals,
combined with the urgency of finding appropriate methodologies to
study defects in HAp and related materials, are the main motivations
behind this work. Below we present a comparative study of different
properties, including the band structure, optical properties and defect
formation energies as function of the Fermi energy, employing three
popular hybrid functionals, namely: the PBE0,\cite{Adamo1999} B3LYP\cite{Stephens1994}
and HSE06.\cite{Krukau2006} The results are compared with quasi-particle
band structure calculations obtained using the highly accurate Green's
function method with screened interactions ($GW$).\cite{Hedin1965,Schilfgaarde2006,Shishkin2007}

\section{Computational details}

\subsection{Semi-local and hybrid density functional calculations}

Calcium apatites have the general chemical formula Ca$_{5}$(PO$_{4}$)$_{3}X$,
with $X$ being an electronegative element such as fluorine, chlorine
or the hydroxyl (OH) group. In the latter case the resulting material
is hydroxyapatite.\cite{Elliott2013} HAp solidifies in the form of
an ionic molecular crystal, either of hexagonal ($P6_{3}/m$) or monoclinic
($P2_{1}/b$) symmetry (\#176 or \#14 in the International Tables
for Crystallography, respectively\cite{Stadnicka1987}), whose unit
cells enclose two or four formula units, respectively.\cite{Kay1964,Elliott1973,Hughes2002}
For the $P6_{3}/m$ phase, hydroxyl units show a stochastic orientation.
Thus, from x-ray data (macroscopic perspective), this effectively
makes the material mirror-symmetric along the main axis. Conversely,
when all OH units show the same alignment along the hexagonal axis,
the mirror plane is lost and the space-group symmetry lowers to $P6_{3}$.

\begin{figure}
\includegraphics[width=1\columnwidth]{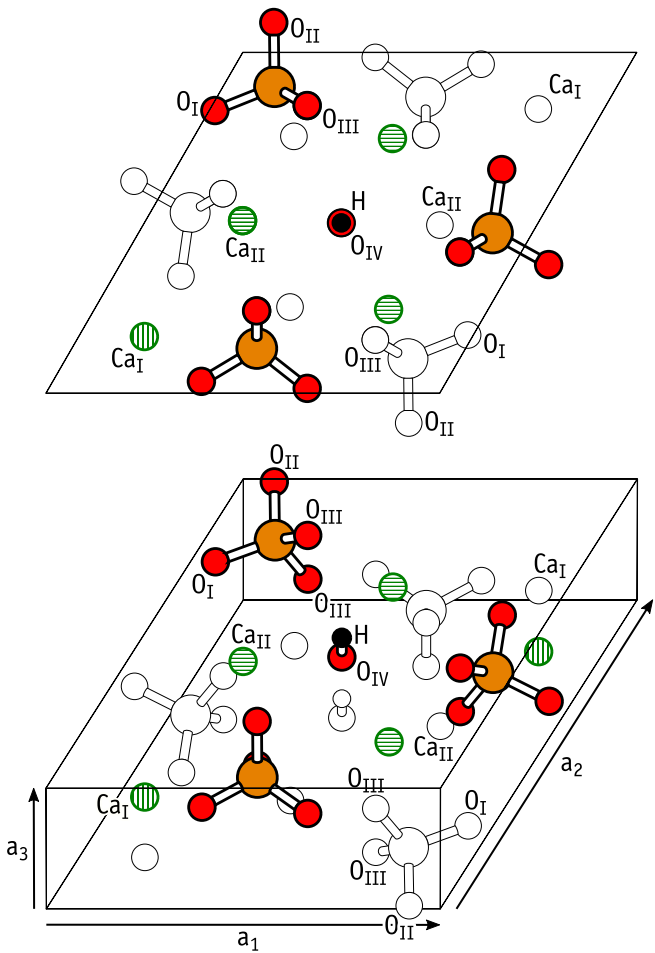}

\caption{\label{fig:struct}Top-projection and three-dimensional representation
of a HAp unit cell centered on the hexagonal channel enclosing the
OH molecular column. Only atoms from one Ca$_{5}$(PO$_{4}$)$_{3}$(OH)
molecular unit are colored. Phosphorus, oxygen calcium and hydrogen
are shown in orange, red, hatched-green and black, respectively. Two
inequivalent Ca atoms and four inequivalent O atoms are distinguished.
Each phosphate molecule has two inequivalent O$_{\mathrm{I}}$ and
O$_{\mathrm{II}}$ sites and two equivalent O$_{\mathrm{III}}$ sites.
Lattice parameters are also indicated.}
\end{figure}

The crystalline structure of HAp ($P6_{3}$ symmetry) is depicted
in Fig.~\ref{fig:struct}, where atoms belonging to one of the formula
units are colored. A total of two and four inequivalent calcium and
oxygen sites are found in HAp, respectively. The Ca$_{\mathrm{I}}^{2+}$
cation columns are surrounded by O$_{\mathrm{I}}$ and O$_{\mathrm{II}}$
from PO$_{4}^{3-}$ anion groups, while mirror-symmetric O$_{\mathrm{III}}$
sites and Ca$_{\mathrm{II}}^{2+}$ ions form a hexagonal channel enclosing
the hydroxyl anions. As mentioned above, different alignments of OH
dipoles in the hydroxyl channels lead to different HAp phases. For
instance, we may have i) a hexagonal disordered phase, with random
orientations of OH dipoles; ii) a hexagonal ordered phase, where OH
dipoles are all oriented along the same direction, or iii) a monoclinic
phase, made of two adjacent cells along a basal lattice vector, with
the first possessing a OH-OH-... column, while the second one showing
an opposite HO-HO-... ordering.\cite{Elliott1973} The last case shows
anti-ferroelectric properties\cite{Ikoma1999} and the symmetry decreases
to monoclinic ($P2_{1}/b$). Given that the electronic band structure
of $P6_{3}$- and $P2_{1}/b$-symmetric HAp are rather similar,\cite{Slepko2011}
the impact of flipping hydroxyl molecules on the properties under
scrutiny is expected to be minor. Hence, we limited our analysis to
the ordered hexagonal phase, allowing us to focus on the effect of
the exchange-correlation treatment.

\begin{figure}
\includegraphics[width=1\linewidth]{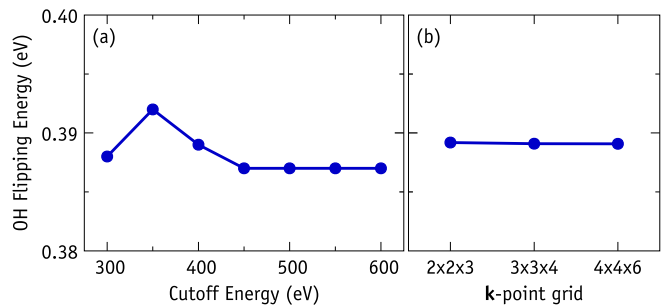}

\caption{\label{fig:conv}Convergence tests showing the energy difference between
two HAp unit cells with OH-OH and OH-HO hydroxyl group alignments
as a function of the plane wave cut-off energy (a), and as a function
of the $\mathbf{k}$-point sampling grid with $E_{\mathrm{cut}}=400$~eV
(b).}
\end{figure}

The calculations were carried out using density functional theory
(VASP package) within the Kohn-Sham formalism.\cite{Kresse1993,Kresse1994,kresse1996a,Kresse1996b}
We employed the projector augmented-wave (PAW) method\cite{Blochl1994}
to account for {[}Ca{]}:1s$^{2}$2s$^{2}$2p$^{6}$3s$^{2}$, {[}P{]}:1s$^{2}$2s$^{2}$2p$^{6}$
and {[}O{]}:1s$^{2}$ core electrons. Valence electrons were described
using plane waves with kinetic energy up to $E_{\mathrm{cut}}=400$~eV.
Convergence tests to the basis quality were carried out by further
increasing $E_{\mathrm{cut}}$ and monitoring the change of the \emph{OH
flipping energy}. This is obtained by subtracting the energy $E_{\mathrm{OH}\textrm{-}\mathrm{OH}}$
of the unit cell shown in Fig.~\ref{fig:struct} (with an OH-OH hydroxyl
alignment) to $E_{\mathrm{OH}\textrm{-}\mathrm{HO}}$ from a cell
with anti-parallel OH-HO alignment. The results obtained within the
GGA to the exchange-correlation are depicted in Fig.~\ref{fig:conv}(a).
For $E_{\mathrm{cut}}=600$~eV the difference $E_{\mathrm{OH}\textrm{-}\mathrm{HO}}-E_{\mathrm{OH}\textrm{-}\mathrm{OH}}$
is 0.387~eV. We also conclude that all $E_{\mathrm{cut}}$ values
give essentially the same answer, which is close to 0.39~eV per unit
cell with an error bar of about 50~meV ($\sim1$~meV/atom) only.
The results are close to 0.4~eV as obtained in Refs.~\onlinecite{Slepko2011}
and \onlinecite{Matsunaga2007} using $E_{\mathrm{cut}}$ values of
700~eV and 500~eV, respectively. They are also compatible with the
experimental estimate for the flipping activation energy of 0.86~eV
(which can be interpreted as approximately $E_{\mathrm{OH}\textrm{-}\mathrm{HO}}-E_{\mathrm{OH}\textrm{-}\mathrm{OH}}$
plus a reverse-flipping barrier).\cite{Yamashita2001} Finally, we
confirmed that increasing $E_{\mathrm{cut}}$ from 400~eV to 600~eV
led to a negligible improvement to the band structure. For instance,
the highest occupied and lowest occupied state energies changed by
less than 8~meV.

The many-body electronic potential was evaluated using either the
GGA according to Perdew, Burke, Ernzerhof (PBE),\cite{perdew1996}
or one of the following hybrid density functionals: (i) The HSE06
of Heyd, Scuseria and Ernzerhof;\cite{Heyd2003,Krukau2006} (ii) The
Becke three-parameter (B3LYP) functional;\cite{Becke1993} (iii) The
PBE0 functional.\cite{Adamo1999} While the exact-exchange contribution
in B3LYP was originally adjusted in order to reproduce experimental
atomic energetics data, for the case of HSE06 and PBE0 functionals,
a fraction $\alpha=1/4$ of Fock exchange is considered, and that
is based on the adiabatic connection formula.\cite{Perdew1996b}

The Brillouin zone (BZ) was sampled using a $\Gamma$-centered 2$\times$2$\times$3
mesh of $\mathbf{k}$-points. Figure~\ref{fig:conv}(b) shows no
noticeable improvements in the quality of the results upon increasing
the sampling density. The Hartree-Fock exact exchange was evaluated
at the same $\mathbf{k}$-point grid used for the DFT potential, and
stored on a real-space grid of 128$\times$128$\times$96 points along
$\mathbf{a}_{1}$, $\mathbf{a}_{2}$ and $\mathbf{a}_{3}$ lattice
vectors, respectively, \emph{c.f.} Fig.~\ref{fig:struct}. The experimental
lattice constants are $a_{1}=a_{2}=a=9.417$~Å and $a_{3}=c=6.875$~Å,\cite{Hughes2002}
corresponding to a grid density of about 14 points/Å along all three
directions.

The calculations of HAp properties were performed on fully relaxed
cells. Lattice relaxations were performed using a quasi-Newton algorithm
combined with an improved basis and real space potential/density grid
($E_{\mathrm{cut}}=520$~eV and a grid comprising 168$\times$168$\times$128
points). The relaxation cycle was stopped when the maximum force acting
on lattice vectors and ions became less than 5~meV/Å.

\subsection{Dielectric response\label{subsec:Dielectric-response}}

The static macroscopic dielectric tensor $\boldsymbol{\epsilon}_{\mathrm{s}}$
couples a time-independent uniform electric field $\boldsymbol{\mathcal{E}}$
to the polarization $\mathbf{P}$ of a material. The polarization
may have different microscopic origins, such as electronic, ionic,
orientational and from space-charge mechanisms. The last two processes
are more important for liquid systems, where polar molecules can rotate
and diffuse rather easier than in solids. Moreover, these mechanisms
are temperature-dependent and are hardly considered within the DFT
framework. Here we consider the electronic polarization within the
adiabatic approximation, as well as polarization effects due to ionic
vibrations (phonons). Although the flipping of OH units are likely
to affect the dielectric response of HAp, such calculations are outside
the scope of the present study. In the following, the dielectric tensor
corresponding to the contribution of the electronic subsystem (\emph{ion-clamped})
is denoted as $\boldsymbol{\epsilon}_{\infty}$, while the ionic contribution
is denoted as $\boldsymbol{\epsilon}_{\text{ph}}$.

The calculation of $\boldsymbol{\epsilon}_{\infty}$ was carried out
using the polarization expansion after discretization (PEAD) approach
of Nunes and Gonze.\cite{Nunes2001} The procedure follows from the
application of a finite homogeneous electric field $\boldsymbol{\mathcal{E}}$
by introducing a set of field-polarized Bloch functions $\{\psi^{\boldsymbol{\mathcal{E}}}\}$
in the Kohn-Sham machinery, and minimizing the electric enthalpy functional,

\begin{equation}
E\left[\left\{ \psi^{\boldsymbol{\mathcal{E}}}\right\} ,\boldsymbol{\mathcal{E}}\right]=E_{0}\left[\left\{ \psi^{\boldsymbol{\mathcal{E}}}\right\} \right]-\Omega\,\boldsymbol{\mathcal{E}}\cdot\mathbf{P}\left[\left\{ \psi^{\boldsymbol{\mathcal{E}}}\right\} \right],
\end{equation}
where $E_{0}$ is the standard (zero-field) density functional, $\mathbf{P}$
is the macroscopic polarization according to the ``modern theory
of polarization''\cite{King-Smith1993} and $\Omega$ is the cell
volume. Accordingly, the ion-clamped macroscopic dielectric tensor
elements are obtained as

\begin{equation}
\epsilon_{\infty,ij}=\delta_{ij}+4\pi\frac{\partial P_{i}}{\partial\mathcal{E}_{j}},
\end{equation}
with $i$ and $j$ representing Cartesian directions. The method accounts
for local-field effects (DFT or hybrid-DFT level) in a natural manner
through the self-consistent procedure. Further details on the PEAD
method and its implementation can be found elsewhere.\cite{Souza2002}

The Born effective charge tensors $\mathbf{Z}_{\alpha}^{*}$ are also
by obtained as a side-product. These couple the applied electric field
to the induced force $\mathbf{F}_{\alpha}$ acting on the $\alpha$-th
atom. Hence,

\begin{equation}
Z_{\alpha,ij}^{*}=\frac{1}{e}\frac{\partial F_{\alpha,j}}{\partial\mathcal{E}_{i}}=\frac{\Omega}{e}\frac{\partial P_{i}}{\partial u_{\alpha,j}},
\end{equation}
is obtained from finite-differences, where $u_{\alpha,j}$ is a sub-lattice
atomic displacement of the $\alpha$-th atom along $j$ and $e$ is
the elementary charge. In our calculations we employed atomic displacements
of 0.015~Å and the magnitude of the electric field was set to 1~meV/Å.
Test calculations within PBE-level showed numerical stability with
electric field magnitudes of up to 10~meV/Å. 

The phonon contribution to the macroscopic dielectric tensor, $\boldsymbol{\epsilon}_{\text{ph}}$,
can be obtained from the Born effective charges and dynamical matrix
elements (second order derivatives of the energy with respect to ionic
positions). The latter was calculated using finite differences with
the atoms being displaced by 0.015~Å along symmetry-irreducible directions.
The eigenvalues $\omega_{\lambda}^{2}$ (with $\lambda=1,\ldots,3N$)
of a dynamical matrix involving $N$ ions correspond to phonon frequencies
$\omega_{\lambda}$ at the BZ center ($\mathbf{q}=0$). Their respective
eigenvectors are the normal modes of vibration $A_{\lambda,\alpha,i}$,
which embody the phonon polarization. According to classical dispersion
theory,\cite{Born1954,Cockayne2000} the $\boldsymbol{\epsilon}_{\text{ph}}$
tensor elements are given by

\begin{equation}
\epsilon_{\mathrm{ph},ij}=\sum\limits _{\omega_{\lambda}^{2}>0}\frac{\tilde{Z}_{\lambda,i}^{\star}\,\tilde{Z}_{\lambda,j}^{\star}}{\Omega\,\omega_{\lambda}^{2}},\label{eq:eps_ph}
\end{equation}
with $\tilde{Z}_{\lambda,i}^{\star}=\sum\limits _{\alpha,j}Z_{\alpha,ij}^{\star}(M_{\alpha})^{-\frac{1}{2}}A_{\lambda,\alpha,j}$
being \emph{vibrational mode effective charges}, where $M_{\alpha}$
is the mass of the $\alpha$-th ion. The phonon contribution to the
dielectric tensor was calculated within PBE-level only.

The static dielectric properties of HAp were measured on crystallite
powder samples,\cite{Kaygili2013,Ikoma1999,Hoepfner2002} so they
effectively behave as macroscopically isotropic. In order to compare
the calculations with the observations we make use of the effective
medium approximation (EMA),\cite{Mayerhofer2002,Mayerhoefer2009}
which defines an orientational-averaged dielectric constant $\epsilon_{\mathrm{s},\mathrm{EMA}}$
as

\begin{equation}
\epsilon_{\mathrm{s,EMA}}=\frac{1}{4}\left(\epsilon_{\mathrm{s},\perp}+\sqrt{\epsilon_{\mathrm{s},\perp}\left(\epsilon_{\mathrm{s},\perp}+8\epsilon_{\mathrm{s},\parallel}\right)}\right),
\end{equation}
with axial $\epsilon_{\mathrm{s},\parallel}=\epsilon_{\mathrm{s},zz}$
and basal $\epsilon_{\mathrm{s},\perp}=\epsilon_{\mathrm{s},xx}=\epsilon_{\mathrm{s},yy}$
elements being parallel and perpendicular to the $c$-axis of the
hexagonal lattice, respectively.

The electronic component of the frequency-dependent macroscopic dielectric
tensor, $\boldsymbol{\epsilon}_{\infty}(\omega)$, characterizes the
response to a long-wave length field (in the limit $\mathbf{q}\rightarrow0$)
with frequency $\omega$. We obtained $\boldsymbol{\epsilon}_{\infty}(\omega)$
within the independent particle picture as derived by Adler and Wiser.\cite{Adler1962,Wiser1963}
Accordingly, the imaginary part of the dielectric tensor, $\boldsymbol{\epsilon}_{\infty}^{(2)}(\omega)$,
is\cite{Gajdos2006}

\begin{multline}
\epsilon_{ij}^{(2)}(\omega)=\frac{4\pi^{2}e^{2}}{\Omega}\lim\limits _{\mathbf{q}\rightarrow0}\frac{1}{q^{2}}\sum\limits _{n,n^{\prime},\mathbf{k}}2w_{\mathbf{k}}\delta(E_{n\mathbf{k}}-E_{n^{\prime}\mathbf{k}}-\hbar\omega)\times\\
\langle u_{n,\mathbf{k}+\mathbf{e}_{i}q}\rvert u_{n^{\prime},\mathbf{k}}\rangle\langle u_{n^{\prime},\mathbf{k}}\rvert u_{n,\mathbf{k}+\mathbf{e}_{j}q}\rangle,\label{eq:imag_eps}
\end{multline}
where $E_{n\mathbf{k}}$ and $E_{n'\mathbf{k}}$ are the Kohn-Sham
eigenvalues of filled and empty states $\psi_{n\mathbf{k}}$ and $\psi_{n'\mathbf{k}}$,
respectively, with wave vector $\mathbf{k}$ within the BZ. The quantities
$\rvert u_{n,\mathbf{k}+\mathbf{e}_{i}q}\rangle$ are obtained using
density-functional perturbation theory,\cite{Gajdos2006} and represent
the first-order change of the cell-periodic part of the wave function
$\psi_{n\mathbf{k}+\mathbf{q}}$ with respect to $\mathbf{q}$, where
$\mathbf{e}_{i}$ are unit vectors for the three Cartesian directions.
Finally, the real part of the dielectric tensor, $\boldsymbol{\epsilon}^{(1)}(\omega)$,
is obtained by the usual Kramers-Kronig transformation.

The number of empty states and $\mathbf{k}$-points in the summation
of Eq.~\ref{eq:imag_eps} should be large enough to produce converged
results. Tests were performed within PBE-level, from which we concluded
that a 4$\times$4$\times$6 $\mathbf{k}$-point mesh and a total
of 400 bands (266 empty states) were enough to reach convergence of
$\boldsymbol{\epsilon}_{\infty}(\omega)$ up to $\omega=30$~eV.

\subsection{Many-body perturbation theory calculations}

We also performed many-body perturbation calculations where the self-energy
was accounted for within the $GW$ approximation (see Refs.~\onlinecite{Shishkin2006,Shishkin2007,Fuchs2007}).
This allowed us to benchmark the description of the HAp band structure,
and by comparison, to assess the performance of DFT and hybrid-DFT
methods. Unlike standard density functional theory, the $GW$ method
accounts for the many-body electron-electron interactions via screening
of the exchange interactions with the frequency-dependent microscopic
dielectric matrix. In practice, quasi-particle energies can be calculated
within the spirit of first-order perturbation theory by solving the
quasi-particle equation

\begin{equation}
E_{n\mathbf{k}}^{\mathrm{QP}}=\mathrm{Re}\left[\left\langle \psi_{n\mathbf{k}}\right|T+V_{\mathrm{ne}}+V_{\mathrm{H}}+\Sigma\left(E_{n\mathbf{k}}\right)\left|\psi_{n\mathbf{k}}\right\rangle \right],\label{eq:qp}
\end{equation}
where $T$ is the kinetic energy operator, $V_{\mathrm{ne}}$ accounts
for the nuclear-electron interactions and $V_{\mathrm{H}}$ is the
Hartree potential. DFT-PBE eigenvalues and wave functions $E_{n\mathbf{k}}$
and $\psi_{n\mathbf{k}}$, respectively, were used in Eq.~\ref{eq:qp}
in order to produce \emph{single-shot} ($G_{0}W_{0}$) quasi-particle
energies. Diagonal elements of the self-energy matrix are given by
\begin{widetext}
\textbf{
\begin{eqnarray}
\left\langle \psi_{n\mathbf{k}}\!\left|\Sigma(\omega)\right|\!\psi_{n\mathbf{k}}\right\rangle  & = & \frac{1}{\Omega}\sum_{\mathbf{q}\mathbf{G}\mathbf{G}'}\sum_{n'}\frac{i}{2\pi}\int_{-\infty}^{+\infty}d\omega'W(\mathbf{G}+\mathbf{q},\mathbf{G}'+\mathbf{q},\omega')\nonumber \\
 &  & \times\frac{\left\langle \psi_{n\mathbf{k}}\left|e^{i(\mathbf{G}+\mathbf{q})\cdot\mathbf{r}}\right|\psi_{n'\mathbf{k}-\mathbf{q}}\right\rangle \left\langle \psi_{n'\mathbf{k}-\mathbf{q}}\left|e^{-i(\mathbf{G}'+\mathbf{q})\cdot\mathbf{r}'}\right|\psi_{n\mathbf{k}}\right\rangle }{\omega+\omega'-E_{n'\mathbf{k}-\mathbf{q}}+i\eta\,\mathrm{sign}(E_{n'\mathbf{k}-\mathbf{q}}-E_{\mathrm{F}})},\label{eq:sigma}
\end{eqnarray}
}where $E_{\mathrm{F}}$ is the Fermi energy and $\eta$ an ubiquitous
but vanishing complex shift. The dynamically screened Coulomb potential,
$W$, was calculated within the random-phase approximation (RPA),
which essentially employs the Hartree potential to account for the
local field effects. The summations in Eq.~\ref{eq:sigma} ran over
a grid of vectors $\mathbf{G}+\mathbf{q}$, whose magnitude was limited
by a cut-off energy $E_{\mathrm{cut}}^{GW}=100$~eV, whereas the
band index $n'$ went up to 1344 bands (1210 of them were empty).
The response function was evaluated considering electronic transitions
between states on a $\Gamma$-centered $4\times4\times6$ $\mathbf{k}$-point
grid (as described above for the macroscopic dielectric tensor), while
the momentum transfer vectors $\mathbf{q}$ were restricted to a coarser
2$\times$2$\times$3 grid. Convergence tests using a denser $4\times4\times6$
$\mathbf{q}$-grid, $E_{\mathrm{cut}}^{GW}=150$~eV and 1410 bands
showed no appreciable changes in the screened Coulomb potential. Finally,
the dielectric function was also calculated from the polarizability
considering the quasi-particle band structure energies.
\end{widetext}

\subsection{Formation energy of defects\label{subsec:defects}}

The procedure for the calculation of formation energies was described
by Qian, Martin and Chadi when investigating the stability of GaAs
surfaces.\cite{Qian1988} Accordingly, the energy needed to introduce
a defect in a crystal (defect formation energy) is $E_{\mathrm{f}}=E-\sum_{i}n_{i}\mu_{i}-\Delta n_{\mathrm{e}}\mu_{\mathrm{e}},$
where $E$ is the total energy of the defective crystal made of $n_{i}$
atomic elements of species $i$ with chemical potential $\mu_{i}$.
This formalism also considers the formation of defects with an excess
of electrons $\Delta n_{\mathrm{e}}$ (with respect to the neutral
state) when they are able to trap/release electrons from/to an electron
reservoir with chemical potential $\mu_{\mathrm{e}}$.

Chemical potentials $\mu_{i}$ can vary within a limited range,

\begin{equation}
\sum n_{i}^{\phi}\left(\mu_{i}-\mu_{i}^{0}\right)\leq\Delta H_{\mathrm{f}}^{\phi},\label{eq:lim_mu}
\end{equation}
with the upper limit taking place when the abundance of species $i$,
pressure and temperature conditions are such that HAp becomes unstable
against segregation of a compound $\phi$ made of $n_{i}^{\phi}$
elements of species $i$ and with heat of formation $\Delta H_{\mathrm{f}}^{\phi}$.
In Eq.~\ref{eq:lim_mu} the quantities $\mu_{i}^{0}$ are chemical
potentials of standard phases (for HAp they stand for the energy per
atom in black phosphorous, molecular oxygen, calcium metal and molecular
hydrogen).

In the calculation of $\mu_{i}^{0}$ related to O$_{2}$ and H$_{2}$
molecules, we could add a term $k_{\mathrm{B}}T\cdot\ln p_{i}^{0}$
in order to account for the partial pressure $p_{i}^{0}$ of the gas
source at temperature $T$.\cite{Matsunaga2007} Here, $k_{\mathrm{B}}$
is the Boltzmann constant. We also note that we are neglecting the
contribution of entropy, although this assumption may be questionable,
especially at high temperatures. Since we are interested in assessing
the effect of the exchange-correlation treatment to the formation
energy of defects, it is actually preferable to leave the results
free of temperature and pressure effects. Note that the location of
the electronic levels within the band gap does not depend on the actual
choice of $\mu_{i}$ values.

Also from Eq.~\ref{eq:lim_mu} we may relate the chemical potentials
$\mu_{i}$ in the HAp solid with the heat of formation per HAp formula
unit as,

\begin{equation}
\sum n_{i}^{\mathrm{HAp}}\left(\mu_{i}-\mu_{i}^{0}\right)=\Delta H_{\mathrm{f}}^{\mathrm{HAp}},\label{eq:heat_fu}
\end{equation}
where the equality sign follows from the equilibrium established between
the HAp solid phase and its own elements. The calculated heat of formation
of HAp per formula unit within PBE and B3LYP levels is $-125.53$~eV
and $-136.19$~eV, respectively. The latter is in close agreement
with $\Delta H_{\mathrm{f}}^{\mathrm{HAp}}=-138.88$~eV as determined
by reaction-solution calorimetry measurements.\cite{Cruz2005}

When using periodic boundary conditions, a more convenient expression
for the defect formation energy is obtained by referring to the number
of atoms $\Delta n_{i}$ of species $i$ added to ($\Delta n_{i}>0$)
or removed from ($\Delta n_{i}<0$) a pristine supercell made of $N^{\mathrm{HAp}}$
formula units with energy $\mu^{\mathrm{HAp}}=\sum_{i}n_{i}^{\mathrm{HAp}}\mu_{i}$,

\begin{equation}
E_{\mathrm{f}}=E-N^{\mathrm{HAp}}\mu^{\mathrm{HAp}}-\sum_{i}\Delta n_{i}\mu_{i}+q(E_{\mathrm{v}}+E_{\mathrm{F}}),\label{eq:formation}
\end{equation}
where $q$ is an integer referring to the charge state of the defect
($q=-\Delta n_{\mathrm{e}}$ is positive/negative when defect levels
within the gap are depleted/filled with $q$ electrons with respect
to the neutral state). The electronic chemical potential is also conveniently
expressed by invoking the Fermi energy so that $\mu_{\mathrm{e}}=E_{\mathrm{v}}+E_{\mathrm{F}}$.
The Fermi energy may vary between the valence band top (where $E_{\mathrm{F}}=0$)
and conduction band bottom ($E_{\mathrm{F}}=E_{\mathrm{g}}$). While
the first and second terms in Eq.~\ref{eq:formation} can be readily
obtained from a first-principles calculation, the last two terms can
vary within certain limits imposed by the thermodynamic conditions.
We also note that the third term only depends on defect-related species.
Here we will look at the formation energy of a OH vacancy, so that
we have to define chemical potentials for O and H only. We assume
that the crystal is in equilibrium with O$_{2}$ and H$_{2}$ molecules
(O- and H-rich conditions), so that $\mu_{\mathrm{O}}=\mu_{\mathrm{O}}^{0}$
and $\mu_{\mathrm{H}}=\mu_{\mathrm{H}}^{0}$. These were calculated
from O$_{2}$ (spin-1 state) and H$_{2}$ molecules in a box of edge
length $L=20$~Å.

\begin{table}
\caption{\label{tab:prop}Experimental (Exp.) and calculated ground state properties
of HAp using different exchange-correlation functionals (PBE, HSE06,
B3LYP and PBE0). Quantities shown are basal and axial lattice constants
($a$ and $c$, respectively), bulk modulus ($B$), pressure derivative
of the bulk modulus ($B'$), ion-clamped ($\epsilon_{\infty}$) and
total ($\epsilon_{\mathrm{s}}$) static dielectric constants along
the direction perpendicular ($\perp$-subscripted) and parallel ($\parallel$-subscripted)
to the $c$-axis. Errors related to $B$ and $B'$ were obtained from
the fitting procedure. The table also shows the the effective medium
approximated dielectric constant $\epsilon_{\mathrm{s,EMA}}$, to
be compared with the orientational-averaged data obtained from powder
samples.}

\begin{ruledtabular}
\begin{tabular}{lccccc}
Property  & Exp.  & PBE  & HSE06  & B3LYP  & PBE0 \tabularnewline
\hline 
$a$ (Å) & 9.417\footnotemark[1] & 9.537 & 9.481 & 9.577 & 9.477\tabularnewline
$c$ (Å) & 6.875\footnotemark[1] & 6.909 & 6.923 & 6.877 & 6.851\tabularnewline
$B$ (GPa)  & $89\pm1$\footnotemark[2]$^,$\footnotemark[3] & $82\pm3$ & $83\pm1$ & $86\pm2$ & $82.8\pm0.3$\tabularnewline
$B^{\prime}$  & $6.9\pm0.6$\footnotemark[3] & $4\pm2$ & $6\pm1$ & $4\pm2$ & $5.4\pm0.3$\tabularnewline
$\epsilon{}_{\infty,\perp}$ & –  & 2.94 & 2.71 & 2.71 & 2.70\tabularnewline
$\epsilon{}_{\infty,\parallel}$ & –  & 2.93 & 2.71 & 2.71 & 2.72\tabularnewline
$\epsilon{}_{\mathrm{s},\perp}$ & –  & 13.1 & 12.8 & 12.8 & 12.8\tabularnewline
$\epsilon{}_{\mathrm{s},\parallel}$  & –  & 8.8 & 8.2 & 8.2 & 8.2\tabularnewline
$\epsilon_{\mathrm{s},\mathrm{EMA}}$  & 7-15\footnotemark[4] & 11.4 & 11.1 & 11.1 & 11.1\tabularnewline
\end{tabular}

\footnotetext[1]{Reference~\cite{Hughes2002}}

\footnotetext[2]{Reference~\cite{Katz1971}}

\footnotetext[3]{Reference~\cite{Gilmore1982}}

\footnotetext[4]{References~\cite{Kaygili2013,Ikoma1999,Hoepfner2002}}
\end{ruledtabular}

\end{table}

Finally, we note that charge neutrality has to be imposed to the whole
cell of any periodic calculation. Hence, the introduction of a localized
charge $q$ is accompanied by a uniform background counter-charge
density $\rho_{\mathrm{back}}=-q/\Omega$.\cite{Makov1995} The Coulomb
interactions between the periodic array of charged defects and background
are long-ranged, and are particularly strong for materials with low
dielectric screening. To mitigate this effect we have to approximate
the \emph{aperiodic} total energy $E$ in Eq.~ \ref{eq:formation}
by a \emph{periodic} total energy plus a correction, $\tilde{E}+E_{\mathrm{corr}}$.
While $\tilde{E}$ is the energy of the defective periodic supercell
as obtained from first-principles, the charge correction was obtained
according to the method proposed by Freysoldt, Neugebauer and Van
de Walle (FNV),\cite{Freysoldt2009} which was recently generalized
for anisotropic materials.\cite{Kumagai2014}

\section{Results and discussion}

\subsection{Ground state properties of HAp\label{subsec:gs-properties}}

We start by reporting on the structural and mechanical properties
of HAp. Table~\ref{tab:prop} compares the calculated unit cell lattice
parameters, bulk modulus, its pressure derivative and static dielectric
properties, with respective experimental data. The table includes
results obtained with PBE, HSE06, B3LYP and PBE0 exchange-correlation
functionals. The calculated lattice parameters within PBE-level are
in line with previous PBE results (see Refs.~\onlinecite{Slepko2011,Zilm2016}
and references therein). They show the usual $\sim1$\% overestimation
with respect to the experimental data. This is known to be mostly
due to an artificial over-delocalization of the electronic density
when the GGA is employed. The B3LYP results improve on $c$ but $a$
is still overestimated by $\sim1$\%. This confirms previous reports
that find B3LYP to generally overestimate the experimental lattice
parameters as well.\cite{Paier2007} On the other hand, our result
differs from previous B3LYP calculations of HAp, where $a$ was underestimated
by 1\%.\cite{Corno2006} This discrepancy is probably due to the use
of a Gaussian-Bloch basis in Ref.~\onlinecite{Corno2006}. Atom-centered
basis functions like these are expected to be less qualified to describe
open-structure solids like HAp, where states overlapping void regions
may be misrepresented. This will become more evident below, when discussing
the electronic structure and screening properties.

Table~\ref{tab:prop} also indicates that the lattice parameters
calculated within hybrid-DFT are generally closer to the experiments
than those obtained using the semi-local functional. The best results
are obtained for PBE0 with a deviation of less than 0.6\% with respect
to the experiments. 

The Bulk modulus ($B$) and its pressure derivative ($B^{\prime}$)
were obtained by fitting the Birch-Murnaghan equation of state,\cite{Birch1947}

\begin{equation}
E(\xi)=E_{0}+\frac{9\Omega_{0}B}{2}\left[\xi^{2}\text{+}\left(B'-4\right)\xi^{3}\right],\label{eq:b-m}
\end{equation}
to total energy data points ($E$) at 8 different cell volumes ($\Omega$)
around equilibrium ($\Omega_{0}$). The volume dependence enters in
Eq,~\ref{eq:b-m} as $\xi=[(\Omega_{0}/\Omega)^{2/3}-1]/2$, and
$E_{0}$ is the minimum energy at $\Omega_{0}$. The calculated bulk
modulus and its pressure derivative show reasonable agreement with
the experiments (see Table~\ref{tab:prop}). The errors reported
on the table were obtained from the fitting procedure. B3LYP calculations
deviates from the measurements by about 3\% only, while other functionals
underestimate the experimental value of $B$ by 7-8\%. This level
of accuracy is in line with typical discrepancies found for many insulators.\cite{Paier2007}

The ion-clamped static dielectric tensors $\boldsymbol{\epsilon}_{\infty}$
as obtained by the different hybrid-DFT schemes (see Table~\ref{tab:prop})
show little variation. They are about 0.2 lower than the PBE results.
This is a property that depends on the ground-state density, and we
interpret this difference as a consequence of the more diffuse PBE
electron density. Interestingly, we find the electronic response of
HAp to be essentially isotropic (within the error bar of the calculation
methodology) with $\epsilon_{\infty,\perp}\approx\epsilon_{\infty,\parallel}\approx2.7$.
Such low anisotropy was also found for strontium apatites within the
LDA.\cite{Yuan2017} Also, the calculated low electronic screening
is consistent with the open-structure and low atomic number of the
HAp constituents.

The ionic contribution $\boldsymbol{\epsilon}_{\mathrm{ph}}$ to $\boldsymbol{\epsilon}_{\mathrm{s}}$
was obtained solely within PBE and we found it considerably larger
and more anisotropic than $\boldsymbol{\epsilon}_{\infty}$. Hence,
the small functional-dependent fluctuations that affect the total
dielectric constants ($\boldsymbol{\epsilon}_{\mathrm{s}}=\boldsymbol{\epsilon}_{\infty}+\boldsymbol{\epsilon}_{\mathrm{ph}}$
in Table~\ref{tab:prop}) derive from variations in $\boldsymbol{\epsilon}_{\infty}$
only. Since the Born effective charges and phonon frequencies depend
solely on the ground state density, not much improvement is anticipated
if a hybrid-DFT method was used. From Eq.~\ref{eq:eps_ph}, we simply
note that the calculation of $\boldsymbol{\epsilon}_{\mathrm{ph}}$
at PBE-level is likely to be affected by the expected underestimation
of the phonon frequencies (which is a characteristic of the PBE functional).

Although the measurement of individual $\epsilon_{\mathrm{s},\perp}$
and $\epsilon_{\mathrm{s},\parallel}$ components for HAp has not
been reported, our values are not far from $\epsilon_{\mathrm{s},\perp}=10.5$
and $\epsilon_{\mathrm{s},\parallel}=8.7$ which were observed for
crystalline fluorapatite.\cite{Shannon1992} The effective medium
approximated dielectric constant is estimated as $\epsilon_{\mathrm{s,EMA}}=11$,
regardless the exchange-correlation functional. The experimental static
dielectric constant determined for HAp powder samples varies widely,
and strongly depends on the porosity and amount of water within the
samples. For example, according to the study of Ikoma \emph{et~al.,}\cite{Ikoma1999}
$\epsilon_{\mathrm{s}}$ varies from 15 to 300 upon heating from 300
to 600~K. On the other hand, Ref.~\onlinecite{Kaygili2013} reports
on the dielectric properties of Ca-doped HAp and arrives at $\epsilon_{\mathrm{s}}=6.75$
only, probably due to the pores in the ceramic pellets. The study
of Hoepfner and Case\cite{Hoepfner2002} addresses precisely the issue
of the porosity dependence of the dielectric constant for sintered
HAp. By extrapolating the data to a vanishing porosity they arrived
at $\epsilon_{\mathrm{s}}\approx14$. This is closer, but somewhat
above the calculated figure, possibly due to contamination of the
samples.

\subsection{Electronic band structure}

\begin{figure*}
\includegraphics[width=1\textwidth]{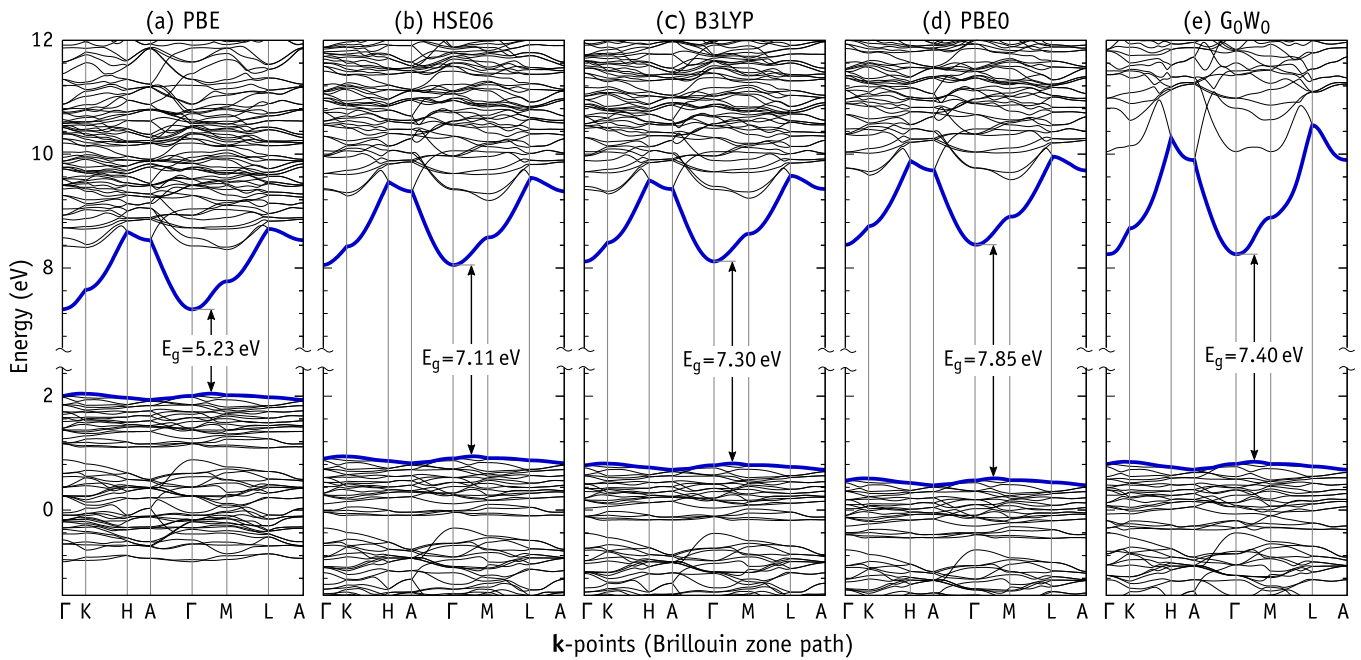}

\caption{\label{fig:band}Electronic band structure of bulk HAp along a path
with kinks at high-symmetry $\mathbf{k}$-points. Calculations were
carried out using DFT within (a) the PBE generalized gradient approximation,
hybrid (b) HSE06, (c) B3LYP and (d) PBE0 functionals, as well as (e)
using the many-body perturbation $G_{0}W_{0}$ method employing the
PBE wave functions. Valence band maximum and conduction band minimum
states are represented as thick lines. Indirect band gap widths, $E_{\mathrm{g}}$,
are also shown for each case.}
\end{figure*}

Figures~\ref{fig:band}(a)-(d) compare the electronic band structure
obtained using different exchange-correlation functionals (PBE, HSE06,
B3LYP and PBE0) with the analogous $G_{0}W_{0}$ quasi-particle calculation
shown in Figure~\ref{fig:band}(e). The band energies along the several
high-symmetry directions were obtained by interpolation of the first-principles
data using Wannier90.\cite{Mostofi2014}

The shape of the PBE band structure in Figure~\ref{fig:band}(a)
is indistinguishable from that reported by Slepko \emph{et~al.}\cite{Slepko2011},
displaying a low-dispersive valence band top and a high-dispersive
conduction band bottom (thick bands). Dispersion of the conduction
band minimum states is considerably more pronounced along directions
parallel to the $c$-axis ($\Gamma\textrm{-}A$, $K\textrm{-}H$,
and $M\textrm{-}L$), indicating a stronger carrier delocalization
and mobility along the main axis. This property could be explored
for tuning HAp electrical conductivity through n-type doping or for
photo-current measurements. On the other hand, p-type doping is not
expected to be beneficial. The valence band top states show very little
dispersion, and their heavy holes imply a relatively lower mobility.

Also in agreement with Ref.~\onlinecite{Slepko2011} we find HAp
to be an indirect-gap material with $E_{\mathrm{g}}=5.23$~eV at
PBE-level. The conduction band minimum is located at $\mathbf{k}=\Gamma$,
while the valence band top energy was found somewhere along $\Gamma\textrm{-}K$
or $\Gamma\textrm{-}M$. The valence band maximum along $\Gamma\textrm{-}M$
is only 0.1~meV higher than the one along $\Gamma\textrm{-}K$. We
note that this picture was the same regardless the functional used,
including when using the $G_{0}W_{0}$ method.

The band structure obtained within HSE06-level is shown in Figure~\ref{fig:band}(b).
The increase in the band gap width by more than 30\% with respect
to the PBE result is self-evident. Using HSE06 we obtain $E_{\mathrm{g}}=7.11$~eV.
Often, the band energies are offset in order to lock the valence band
top at the origin of the energy scale. We did not follow this procedure,
and that allowed us to disclose how the gap change depends on the
shift of both valence band and conduction band states. Figures~\ref{fig:band}(b)-(d)
show that admixing a fraction of Fock exchange with the semi-local
exchange energy has a profound effect on both valence and conductions
band states. Consequently, the use of hybrid functionals has implications
not only to the accuracy of calculated defect-related or inter-band
transitions (\emph{e.g.} observed in luminescence or UV-VIS absorption),
but also to transitions involving core or vacuum states (\emph{e.g.}
observed in electron photoemission or core electron energy loss spectroscopy).

The gap of the B3LYP band structure depicted in Figure~\ref{fig:band}(c)
is 0.6~eV narrower than that reported by Corno \emph{et~al.}\cite{Corno2006}
using the same functional. Again, the difference is likely to be due
to the unsuitability of the atomic-like basis employed in Ref.~\onlinecite{Corno2006},
which resulted in the under-screening of the band structure. The band
gap width as well as the band gap edge energies obtained at the B3LYP-level
are closer to the $G_{0}W_{0}$ results than any other functional.

The trend of the calculated band gaps agrees with that obtained by
Garza and Scuseria\cite{Garza2006} for an eclectic mix of semiconductors
and insulators. Accordingly, HSE06 and B3LYP showed a closer correlation
with the experiments (the former giving slightly smaller gaps overall),
whereas inclusion of larger fractions of exact exchange like in PBE0,
led to an overestimation of $E_{\mathrm{g}}$. This ordering also
suggests that the gap width obtained within $G_{0}W_{0}$ ($E_{\mathrm{g}}=7.4$~eV)
should be close to the real figure. It is also interesting to note
that the frequency-dependent screening of the electronic interactions
within $G_{0}W_{0}$ has a larger impact on empty states, where several
near-degeneracies obtained at DFT level, become well separated when
using the many-body perturbation method (see Figure~\ref{fig:band}(e)).

\begin{figure}
\includegraphics[width=0.9\columnwidth]{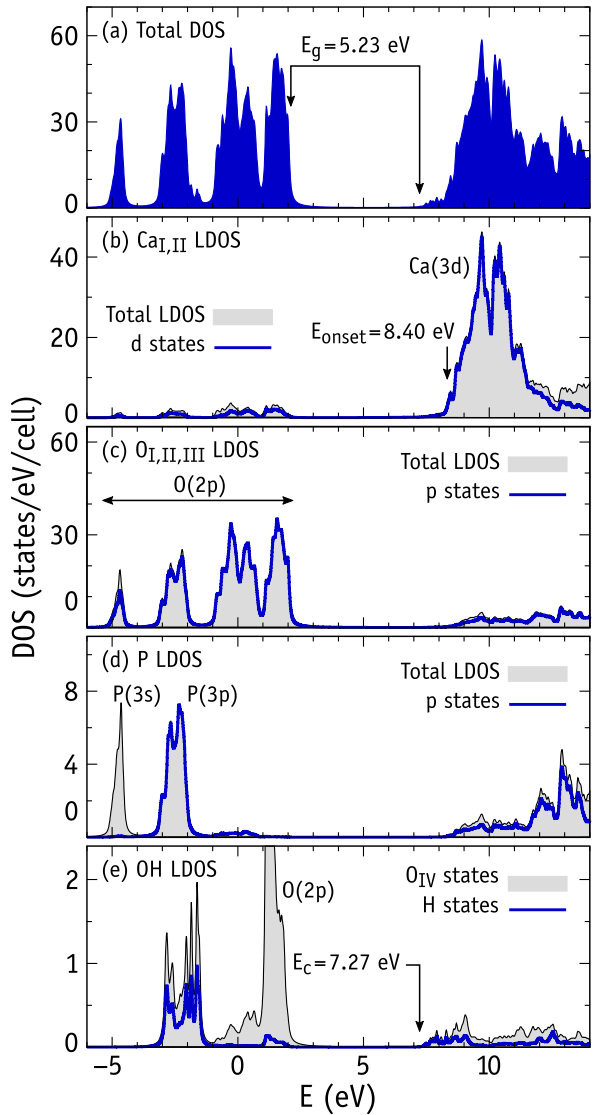}

\caption{\label{fig:pdos}Total density of states of bulk HAp (a) and local
densities of states projected on calcium (b), oxygen in PO$_{4}$
units (c), phosphorous (d), as well as oxygen and hydrogen in OH units
(e). In (b)-(d) the contribution to the LDOS with the dominant angular
momentum is represented as a thick line. The results were obtained
using the PBE exchange-correlation functional.}
\end{figure}

Figure~\ref{fig:pdos}(a) shows the total density of states (DOS)
of a HAp unit cell obtained within PBE. The energy scale is directly
comparable to the band structure of Figure~\ref{fig:band}(a). The
shadow plots on subsequent Figs.~\ref{fig:pdos}(b)-(e), depict the
local density of states (LDOS) projected on several atomic species.
Along with the LDOS of Figs.~\ref{fig:pdos}(b)-(d), we plot a thick
line representing the dominant angular-momentum component for the
corresponding species. In Fig.~\ref{fig:pdos}(e) we distinguish
states projected on O$_{\mathrm{IV}}$ and H atoms that form OH molecules.
The LDOS calculations considered a Wigner-Seitz projection radius
of 1.75~Å, 1.23~Å, 0.82~Å and 0.37~Å for Ca, P, O and H respectively.
For the sake of clearness, the data was convoluted using 0.1~eV wide
Lorentzian functions. 

At first glance, Figs.~\ref{fig:pdos}(b) and \ref{fig:pdos}(c)
suggest that the upper end of the valence band is mostly made of O(2p)
states, while the conduction band bottom is mostly made of Ca(3d)
states. From Figs.~\ref{fig:pdos}(c) and \ref{fig:pdos}(d) we find
phosphorous 3s-3p states mixing with oxygen 2s-2p states between $-5$
and $-2$~eV, and they are far below from the band gap region. These
results are in line with previous reports.\cite{Slepko2011,Matsunaga2007}

Regarding the origin of the bottom of the conduction band, in particular
the highly dispersive bands shown in Fig.~\ref{fig:band}, the situation
is more controversial. In Ref.~\onlinecite{Slepko2011}, a close
analysis of the LDOS at the conduction bottom indicated that the lowest
energy bands originated from Ca(4s) states. We argue that this view
finds several difficulties. Firstly because we find the onset of the
Ca-LDOS ($E_{\mathrm{onset}}=8.40$~eV) located $\sim\!1$~eV above
the conduction band minimum energy. Within that energy range, all
that can be related to Ca are the flat Ca(3d) bands above 8.4~eV.
Secondly, Figure~\ref{fig:pdos}(e) shows that states just above
$E_{\mathrm{c}}$ have a considerable localization on OH molecules.
Interestingly, we found that the highly dispersive bands that form
the bottom of the conduction band of HAp, are anti-bonding states
from an infinite $\cdots\textrm{OH-OH-}\cdots$ hydrogen bridge sequence,
much like a 1D-ice phase. This statement finds support in the band
structure of hexagonal ice (see for example Figures~4 and 6 of Refs.~\onlinecite{Prendergast2005}
and \onlinecite{Engel2015}, respectively). From comparison with our
Figure~\ref{fig:band}(a), it becomes immediately evident that the
dispersion shape of the lowest conduction band of HAp, is analogous
to that of hexagonal ice.

\begin{figure}
\includegraphics[width=1\columnwidth]{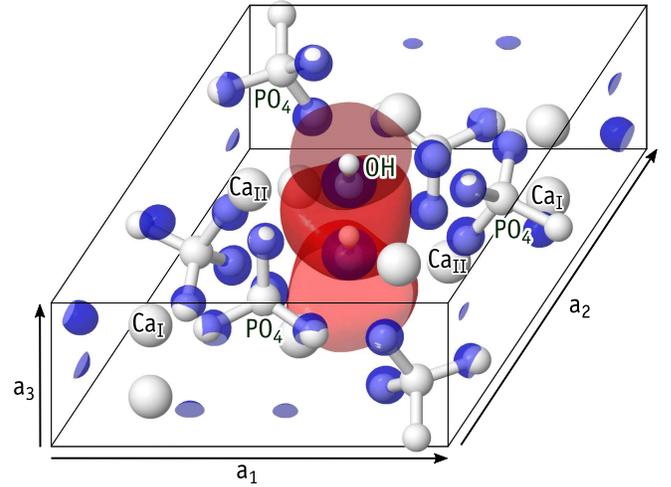}

\caption{\label{fig:wf}Lowest unoccupied Kohn-Sham state (bottom of the conduction
band) of a HAp at $\mathbf{k}=\Gamma$. Blue and red isosurfaces represent
$\psi(\mathbf{r})=+0.02$ and $\psi(\mathbf{r})=-0.02$ phases of
the wave function. All atoms are shown in white.}
\end{figure}

Finally, our view is demonstrated by Figure~\ref{fig:wf}, which
depicts the lowest unoccupied Kohn-Sham state of a HAp at $\mathbf{k}=\Gamma$,
using $\psi(\mathbf{r})=\pm0.02$ isosurfaces. Positive and negative
phases are shown in blue and red, respectively. It is unquestionable
that the conduction band bottom of HAp is made of anti-bonding $\sigma^{*}$
states along the OH linear chain, with a minor localization on oxygen
atoms of PO$_{4}$ units, and even less on Ca. The diffuse nature
of this state makes it difficult to be projected into atomic-like
orbitals, and that could be the reason behind the misassignment in
Ref.~\onlinecite{Slepko2011}. This finding may have important consequences
for doping strategies. For instance, among the possible atomic sites
for doping, the replacement of calcium by a foreign species is likely
to have the least disruptive effect on the conductive properties of
the OH chain.

\subsection{Optical properties}

\begin{figure}
\includegraphics[width=0.9\linewidth]{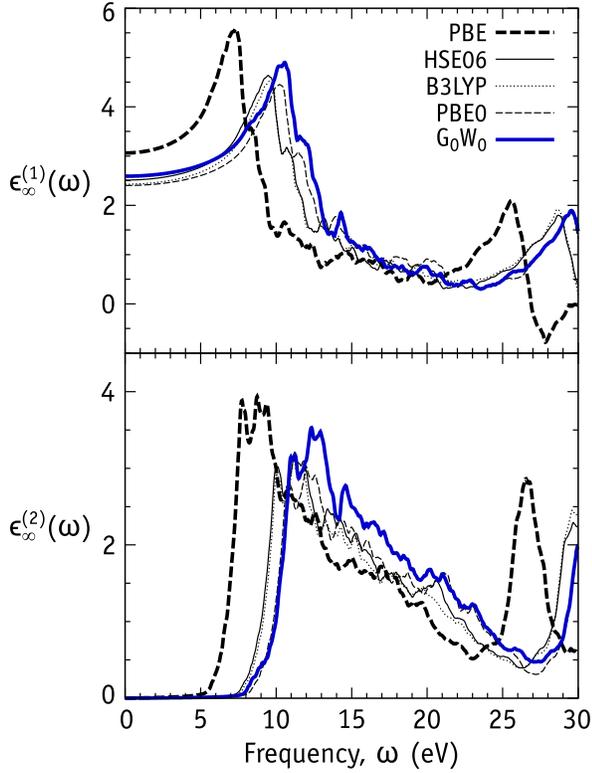} \caption{\label{fig:eps}(a) Real part $\epsilon_{\infty}^{(1)}$ and (b) imaginary
part $\epsilon_{\infty}^{(2)}$ of the ion-clamped frequency-dependent
dielectric function obtained using PBE, HSE06, B3LYP, PBE0 functionals,
as well as using the $G_{0}W_{0}$ quasi-particle energies.}
\end{figure}

Figure~\ref{fig:eps} shows the real and imaginary parts of the ion-clamped
dielectric function calculated using density functional perturbation
theory. For the sake of clearness, we only report the parallel component
of $\epsilon_{\infty}$. The basal component was essentially identical
to the axial component. The $G_{0}W_{0}$ response curves were obtained
employing the quasi-particle corrected band structure.

The shape of the dielectric function obtained using the PBE exchange-correlation
functional (tick dashed line) is in general close to that reported
by Rulis \emph{et~al.}\cite{Rulis2004} obtained within the LDA.
The imaginary part, $\epsilon_{\infty}^{(2)}$, is proportional to
the absorption coefficient due to inter-band transitions. This allows
us to obtain some details of $\epsilon_{\infty}(\omega)$ from inspection
of the band structure and DOS/LDOS plots of Figures~\ref{fig:band}
and \ref{fig:pdos}. Accordingly, the onset of $\epsilon_{\infty}^{(2)}$
takes place at about 5.2~eV, consistent with the PBE band gap, and
therefore to the fundamental transition from the valence band top
and the OH-related conduction band bottom. The electronic contribution
to the static dielectric constant is $\epsilon_{\infty}^{(1)}(\omega\rightarrow0)\approx3$,
consistent with $\epsilon_{\infty}=2.93$ as obtained from the calculated
electronic polarization upon application of a constant electric field
(see Table~\ref{tab:prop}).

The first major pole in the dielectric function is located at about
8~eV. It results from transitions from the top-most O(2p) states
shown between 1-2~eV in the DOS plot, to the Ca(3d)-related bands.
The structure displayed in $\epsilon_{\infty}^{(2)}$ below 15~eV
arises from transitions involving the several O(2p) bands in the valence
to the broad Ca(3d) DOS (see also Fig.~\ref{fig:pdos}). The second
prominent pole at about 27~eV arises from internal Ca(3p)$\rightarrow$Ca(3d)
transitions.

Regarding the results obtained with the hybrid functionals, three
main differences are seen when compared to PBE. First, the static
dielectric constant using the hybrid functionals decreases by about
0.5. Within hybrid-DFT we obtain $\epsilon_{\infty}^{(1)}(\omega\rightarrow0)\approx2.5$,
close to $\epsilon_{\infty}=2.7$ from the PEAD/hybrid-DFT method
in Section~\ref{subsec:gs-properties}, but below $\epsilon_{\infty}^{(1)}(\omega\rightarrow0)\approx3$
using the PBE approximation. Secondly, all hybrid functionals lead
to a blue shift of the dielectric function by 2-3~eV, consistent
with an increase of $E_{\mathrm{g}}$ from 5.23~eV to about 7.5 eV.
The magnitude of this effect is about the same for all hybrid functionals,
although slightly more pronounced for PBE0. In fact, PBE0 leads to
results closer to those that use the quasi-particle band structure
from $G_{0}W_{0}$ calculations. Finally, both real and imaginary
parts of the dielectric function show a lower amplitude when using
hybrid functionals and the quasi-particle band structure. This follows
from the relatively less ``crowded'' conduction bands when compared
to those obtained with PBE (see Figure~\ref{fig:band}).

\subsection{Defect formation energies: The OH-vacancy}

We now turn to the assessment of the exchange-correlation treatment
in the calculation of defect formation energies and their electronic
levels. This was done by looking at a simple point defect, namely
the OH-vacancy (hereafter referred to as V$_{\mathrm{OH}}$). We used
352-atom HAp supercells by replicating $2\times2\times2=8$ hexagonal
unit cells, and removing a single hydroxyl unit to create a V$_{\mathrm{OH}}$
defect. By doubling the size of the cell along all principal directions,
the reciprocal lattice vectors are contracted by a factor of two,
so that a $1\!\times\!1\!\times\!2$ grid of $\mathbf{k}$-points
(to construct the density) would even provide us better sampling quality
than the $2\!\times\!2\!\times\!3$ grid used for the primitive cell.
Convergence tests showed that the total energy of the 352-atom bulk
supercell with $\Gamma$-point sampling only, differs by less than
0.1~eV (0.3~meV/atom) from a $1\!\times\!1\!\times\!2$-sampled
calculation. Therefore we employed a $\Gamma$-point sampling for
defective supercells.

Due to the dramatic impact of the exact exchange treatment in the
band structure, the use of hybrid functionals should have important
repercussions on the calculated formation energies, particularly on
the accuracy of the calculated electronic transition levels. From
a thermodynamic perspective, a $E(q/q\!+\!1)$ transition level is
the location of the Fermi energy within the gap, for which equal populations
of defects in charge states $q$ and $q+1$ are found in equilibrium.
This takes place when formation energies of both charge states (calculated
with help of Eq.~\ref{eq:formation}) are equal for a particular
value of $E_{\mathrm{F}}$.

\begin{figure}
\includegraphics[width=0.95\linewidth]{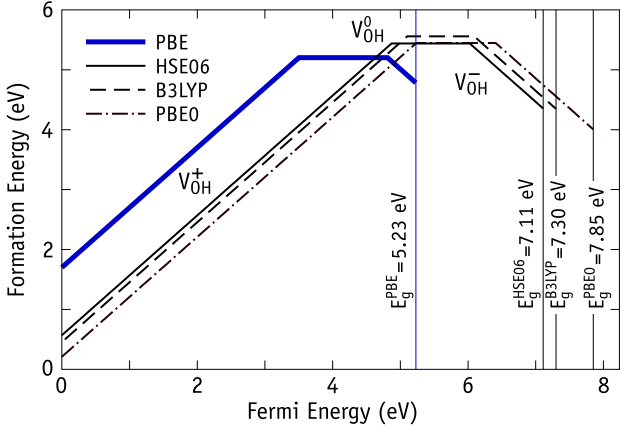} \caption{\label{fig:Evac}Calculated formation energy of V$_{\mathrm{OH}}$
in HAp as a function of the Fermi energy, using different approximations
to the exchange-correlation interactions. Positive, neutral and negative
charge states are represented by lines with positive, zero and negative
slope, respectively. For each functional, the Fermi energy can vary
between the valence band top (at $E_{\mathrm{F}}=0$~eV) and the
respective conduction band minimum ($E_{\mathrm{F}}=E_{\mathrm{g}}$).
Chemical potentials of O and H were obtained from O$_{2}$ and H$_{2}$
molecules. }
\end{figure}

We found that the band structure of HAp with a V$_{\mathrm{OH}}$
defect differs from that of bulk, only by the presence of an additional
semi-occupied defect-related band close to mid-gap. This suggests
that V$_{\mathrm{OH}}$ can donate an electron (leaving the defect
band empty) or accept and electron (filling up the defect band). Figure~\ref{fig:Evac}
depicts the formation energy of V$_{\mathrm{OH}}^{q}$ as a function
of the Fermi energy, where $q$ is the defect charge state. Positive-,
zero- and negative-slope lines represent formation energies of positively
charged, neutral and negatively charged vacancies (V$_{\mathrm{OH}}^{+}$,
V$_{\mathrm{OH}}^{0}$ and V$_{\mathrm{OH}}^{-}$), respectively.
The thick solid lines are the results obtained within PBE-level, while
the thin lines represent formation energies obtained using the hybrid
functionals (HSE06, B3LYP and PBE0). For each case, the Fermi energy
can vary between $E_{\mathrm{F}}=0$ (for $p$-type material) and
$E=E_{\mathrm{g}}$ (for $n$-type material), with $E_{\mathrm{g}}$
being the respective band gap width.

The values of $E_{\mathrm{F}}$ at the ``kinks'' joining the line
segments define the $E(q/q\!+\!1)$ transitions. Hence, within PBE
we obtain donor and acceptor transitions at $E(0/+)=E_{\mathrm{v}}+3.50$~eV
and $E(-/0)=E_{\mathrm{v}}+4.81$~eV, respectively. On the other
hand, using the hybrid functionals we have donor and acceptor transitions
at $E(0/+)=E_{\mathrm{v}}+(5.1\pm0.2)$~eV and $E(-/0)=E_{\mathrm{v}}+(6.2\pm0.3)$~eV,
respectively. The upper and lower limits are obtained using HSE06
and PBE0, respectively. The average values are close to the B3LYP
results and differ from the PBE results by about 1.5~eV. This is
considerably less than the corresponding change in $E_{\mathrm{g}}$,
suggesting that the error of (semi)local-functional calculations of
electronic transitions, depends much on the character of the defect
state that is being populated or emptied. It is also interesting to
note that the formation energy of neutral V$_{\mathrm{OH}}$ using
the hybrid functionals is only 0.2~eV above the PBE result. However,
there is no reason to conclude that we have found a generality for
neutral defects. We assume that this small difference was simply fortuitous.

\subsection{Discussion on corrections to charged defects}

As noted in the last paragraph of Section~\ref{subsec:defects},
total energies of charged defects were subject a charge correction
$E_{\mathrm{corr}}$ to remove the effect of artificial Coulomb interactions
between periodic charged replicas. A question that arises during the
calculation of $E_{\mathrm{corr}}$ is — should we use the electronic
screening constant $\epsilon_{\infty}$ or the total static dielectric
constant $\epsilon_{\mathrm{s}}$? As suggested by Komsa $et~al.$\cite{Komsa2012},
for the calculations of unrelaxed defects (where ligand atoms occupy
perfect crystallographic sites), it is appropriate to use the high-frequency
(ion-clamped) dielectric constant for describing the screening. However,
when atoms are allowed to relax and polarize the supercell, the static
low-frequency dielectric constant ($\epsilon_{\mathrm{s}}$) should
be employed instead.

\begin{figure}
\includegraphics[width=1\linewidth]{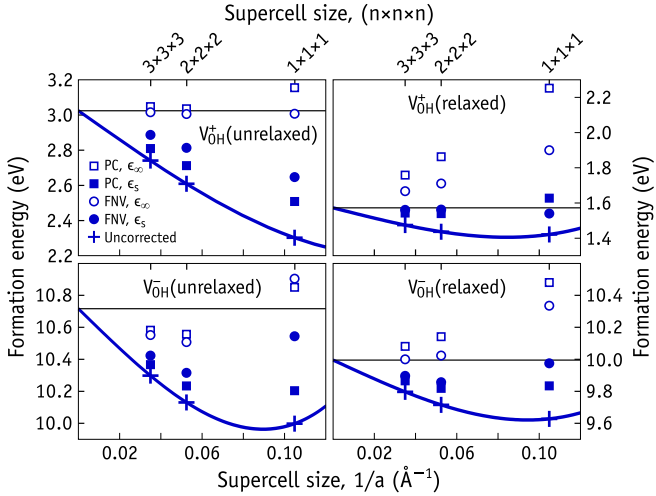} \caption{\label{fig:corr} Formation energy of V$_{\mathrm{OH}}^{+}$ and V$_{\mathrm{OH}}^{-}$
in relaxed and unrelaxed supercells, assuming $E_{\mathrm{F}}=0$.
Different charge correction schemes are compared on each plot, namely
without any charge correction (crosses), with a simple point-charge
correction (squares), and with the FNV correction scheme. Results
obtained with the ion-clamped and total static dielectric constants
are shown as open and solid symbols. Data in the figure were calculated
within PBE level. See text for further details.}
\end{figure}

The above can be verified with help of Figure~\ref{fig:corr}, were
we plot the formation energy of both fully relaxed and unrelaxed V$_{\mathrm{OH}}^{+}$
and V$_{\mathrm{OH}}^{-}$ defects, considering $E_{\mathrm{F}}=0$.
All data were obtained within PBE level. Tests on HAp unitcells and
$2\times2\times2$ supercells using B3LYP indicated that the dependence
on the particular exchange-correlation functional was negligible.
Each plot contains either (i) uncorrected formation energies (shown
as crosses), (ii) formation energies corrected by the simple point-charge
scheme proposed by Makov and Payne\cite{Makov1995} (shown as squares),
and (iii) formation energies corrected by using the FNV method.\cite{Freysoldt2009,Kumagai2014}
The effect of the screening constant on the results can be examined
from the difference obtained using either the ion-clamped $\epsilon=\epsilon_{\infty}\approx3$,
or the total static dielectric constant $\epsilon=\epsilon_{\mathrm{s}}\approx11$
(open and solid symbols, respectively). The results are shown as a
function of the inverse of the basal dimension of the supercell (lower
horizontal axis) and as function of the number of unit cells replicated
along each lattice direction (upper horizontal axis). The largest
supercell employed had 1188 atoms in bulk, being obtained by replication
$3\times3\times3$ of unit cells along all crystallographic directions.
The uncorrected data is used to expand the formation energy in a power
series $E(a)=E_{0}+E_{1}a^{-1}+E_{3}a^{-3}$ as proposed in Ref.~\onlinecite{Castleton2006}.
The second and third terms account for the point-charge-like and dipole-like
dependence of the Coulomb correction, respectively. The fit allows
us to obtain the formation energy extrapolated to an infinite cell
where the charge correction vanishes (horizontal thin line).

It is clear that the correction to be made strongly depends on the
defect and its charge state. For the $2\times2\times2$ supercells
used to obtain the data of Figure~\ref{fig:Evac} we have $E_{\mathrm{corr}}$
values of about 0.2~eV and 0.3~eV for positively and negatively
charged defects. Although they are a small fraction of $E_{\mathrm{g}}$,
they are expected to be 4 times larger for double positive or double
negative charge states. This is because the leading term of the correction
scales as $E_{\mathrm{corr}}\sim q^{2}$. Another feature, which is
well known in the literature (see for instance Refs.~\onlinecite{Freysoldt2009}
and \onlinecite{Kumagai2014}), is the fact that the simple point-charge
scheme tends to over-estimate the correction. In that respect, the
FNV method performs better.

Comparing the plots on the left (unrelaxed) and on the right (relaxed)
of Figure~\ref{fig:corr}, we realize that for the unrelaxed cells,
the methods that employ the ion-clamped dielectric constant ($\epsilon_{\infty}$)
are more accurate than using the static dielectric constant ($\epsilon_{\mathrm{s}}$).
On the other hand, for the relaxed cells we obtain an opposite behavior.
This shows that the calculation of defect formation energies which
involve the relaxation of the atomic positions, should employ the
static dielectric constant for the evaluation of the charge correction.

\section{Conclusions}

We report on the impact of the exchange-correlation treatment to the
calculation of several properties of pristine and defective hydroxyapatite.
The work was carried out using density functional theory, employing
the PBE scheme for the GGA, as well as three popular hybrid-functionals
which mix the semi-local exchange with exact Fock exchange. The hybrid-functionals
considered are the HSE06, B3LYP and PBE0, and their performance is
compared to state-of-the-art many-body perturbation calculations,
where the self-energy is accounted for within the $GW$ approximation.

We start by discussing how well the different exchange-correlation
treatments describe the fundamental structural and mechanical properties
of HAp. We found that all hybrid-functionals outperform the PBE method,
but still, they overestimate the experimental lattice parameters by
about 1\%, while bulk modulus are underestimated by 3-6\%.

The static dielectric response of HAp was investigated by the polarization
expansion after discretization method. The electronic contribution
to the static dielectric constant was found approximately isotropic.
A value of $\epsilon_{\infty}=2.7$ was obtained, irrespectively of
the choice of the hybrid-functional. This overestimates the PBE value
by 0.2 only, suggesting that the semi-local exchange treatment already
provides a reasonable description of the ground state density. We
also show that most screening and anisotropy effects are related to
the polarization of the lattice by phonons. The components of the
static dielectric constant along the basal plane and along the main
crystallographic axis, are anticipated as $\epsilon_{\mathrm{s},\perp}=12.8$
and \foreignlanguage{english}{$\epsilon_{\mathrm{s},\parallel}=8.2$}.
According to the effective medium approximation, these figures correspond
to an isotropic dielectric constant $\epsilon_{\mathrm{EMA}}=11$,
which is within the observed range of 6-14 from powder samples.

The electronic band structure within PBE is in line with previous
equivalent calculations. While the valence band top is made of highly
localized O(2p) states, the conduction band bottom of HAp is delocalized
and shows strong dispersion. These overlap the periodic $\cdots\textrm{OH-OH-}\cdots$
hydrogen bridge chain and carry an anti-bonding $\sigma^{*}$ character,
suggesting the formation of a 1D-ice phase. This property could be
used to explore electrical transport in n-type doped HAp, or for applications
involving photo-current. The use of hybrid-DFT was found to have a
huge impact on both valence band and conduction band energies. It
led to an increase of the band gap width by more than 30\% when compared
to PBE results (5.23~eV). The $GW$ quasi-particle band gap was 7.4~eV
wide, and the closest hybrid-DFT figure was obtained within B3LYP
(7.3~eV). We conclude that (semi-)local calculations of many properties
such as defect-related or inter-band transitions (defect levels, UV-VIS
spectra), as well as transitions involving core or vacuum states (energy
loss or photo-emission spectroscopy) can be largely improved by mixing
a portion of exact exchange into the density functional.

Regarding the optical properties, the dielectric function obtained
from hybrid-DFT and $GW$ band structure was blue-shifted with respect
to the PBE result by 2-3~eV. This is consistent with the increase
of $E_{\mathrm{g}}$ from 5.23~eV to about 7.5~eV using PBE and
hybrid-DFT/$GW$ methods, respectively. Both real and imaginary parts
of the dielectric function show a lower amplitude when using hybrid
functionals and the quasi-particle band structure. The PBE0 leads
to results closer to the $GW$ calculations.

The choice of the exchange-correlation functional and its impact on
the calculation of defect levels was also investigated. We used the
OH-vacancy as a testing-model to look at donor and acceptor levels.
The levels obtained within hybrid-DFT were markedly different from
the PBE results and the difference was not the same for all levels.
We finally conclude that any accurate estimation of defect levels
using HAp supercells, will require the introduction of a charge correction.
This removes spurious interactions between periodic charged replicas
from the total energy. The methods available invariably make use of
a dielectric constant. It is emphasized that atomic relaxations around
the defect polarize the material beyond the ion-clamped level. Hence,
the static dielectric constant to be used must include both electronic
and ionic polarization effects.

\section*{Acknowledgements}

This work was supported by the \emph{Fundação para a Ciência e a Tecnologia}
(FCT) through project UID/CTM/50025/2013 and by the Russian Foundation
for Basic Research (RFBR) through project No.~15-01-04924. SÖ acknowledges
the Swedish National Infrastructure for Computing (SNIC) at PDC for
providing computational resources. 

\bibliographystyle{apsrev4-1}

%

\end{document}